\documentclass[prl,twocolumn,superscriptaddress,floatfix,nobalancelastpage]{revtex4}

\usepackage{amsmath}
\usepackage{amssymb}
\usepackage{graphicx}
\usepackage{dcolumn}

\begin{document}

\title{Fermi surface of SrFe$_2$P$_2$ determined by de Haas-van Alphen effect}

\author{J.G. Analytis}
\affiliation{Stanford Institute for Materials and Energy Sciences, SLAC National Accelerator Laboratory, 2575 Sand Hill
Road, Menlo Park, CA 94025, USA} \affiliation{Geballe Laboratory for Advanced Materials and Department of Applied
Physics, Stanford University, USA}

\author{C.M.J. Andrew}
\affiliation{H.H. Wills Physics Laboratory, University of Bristol, Tyndall Avenue, Bristol, BS8 1TL, UK}

\author{A.I. Coldea}
\affiliation{H.H. Wills Physics Laboratory, University of Bristol, Tyndall Avenue, Bristol, BS8 1TL, UK}

\author{A. McCollam}
\affiliation{Radboud University Nijmegen,High Field Magnet Laboratory, Faculty of Science,6500 GL Nijmegen, The
Netherlands.}

\author{J.-H. Chu}
\affiliation{Stanford Institute for Materials and Energy Sciences, SLAC National Accelerator Laboratory, 2575 Sand Hill
Road, Menlo Park, CA 94025, USA} \affiliation{Geballe Laboratory for Advanced Materials and Department of Applied
Physics, Stanford University, USA}

\author{R.D. McDonald}
\affiliation{Los Alamos National Laboratory, Los Alamos, NM 87545, USA}

\author{I.R. Fisher}
\affiliation{Stanford Institute for Materials and Energy Sciences, SLAC National Accelerator Laboratory, 2575 Sand Hill
Road, Menlo Park, CA 94025, USA} \affiliation{Geballe Laboratory for Advanced Materials and Department of Applied
Physics, Stanford University, USA}

\author{A. Carrington}
\affiliation{H.H. Wills Physics Laboratory, University of Bristol, Tyndall Avenue, Bristol, BS8 1TL, UK}

\begin{abstract}
We report measurements of the Fermi surface (FS) of the ternary iron-phosphide SrFe$_2$P$_2$ using the de Haas-van
Alphen effect. The calculated FS of this compound is very similar to SrFe$_2$As$_2$, the parent compound of the high
temperature superconductors.  Our data show that the Fermi surface is composed of two electron and two hole sheets in
agreement with bandstructure calculations. Several of the sheets show strong $c$-axis warping emphasizing the
importance of three-dimensionality in the non-magnetic state of the ternary pnictides. We find that the electron and
hole pockets have a different topology, implying that this material does not satisfy a ($\pi,\pi$) nesting condition.
\end{abstract}


\maketitle

Many theories of the superconductivity and magnetism in the iron-pnictides have their unusual Fermi surface topology as
a central ingredient \cite{Chubukov2008a,Mazin2008a,stanev_spin_2008,ran_nodal_2009}.  Both superconductivity and
magnetism can be enhanced by geometrical nesting of the hole and electron Fermi surface sheets. Bandstructure calculations
suggest that 1111-arsenides (ReFeAsO$_{1-x}$F$_x$, Re $\in$ La$\rightarrow$Sm) have stronger `nesting' peaks in the
non-interacting susceptibility than their 122 arsenide (XFe$_2$As$_2$, X $\in$ Eu, Ba, Sr, Ca) counterparts and the
former generally have a higher $T_c$.  The 122 materials are calculated to have stronger warping of their
quasi-two-dimensional sheets and experimentally this is reflected in a more isotropic upper critical field $H_{c2}$.

One argument against a predominant role of Fermi surface nesting is that the phosphide analogues of the 122 arsenide
materials (XFe$_2$P$_2$) do not show superconducting or magnetic order despite their calculated Fermi surfaces being
very similar to the corresponding arsenides.  However, partial substitution of As by P in EuFe$_2$(As$_{1-x}$P$_x$)$_2$
\cite{ren_xu} and BaFe$_2$(As$_{1-x}$P$_x$)$_2$ \cite{jiang_xu} suppresses the antiferromagnetic order and results in
superconductivity with $T_c$ up to 26\,K and 30\,K respectively.  The evolution from antiferromagetism to
superconductivity and then to paramagnetism in this series provides a strong test of our understanding of the physics
of these materials.  A particular question is whether this is driven by changes in the Fermi surface topology. Although
bandstructure calculations can provide a guide, accurate experimental determinations are essential because inaccuracies
in the calculated band energies (often of order 0.1\,eV \cite{coldea}) can lead to significant changes in the topology.

Quantum oscillation (QO) studies prove a direct way to probe in detail the bulk, full three dimensional Fermi surface,
and can also resolve the strength of the many-body interactions at the Fermi level. However, so far QO experiments have
not been possible on the high $T_c$ superconducting arsenides because of their high $H_{c2}$ and disorder induced by
doping.  Here we present a de Haas-van Alphen effect determination of the Fermi surface topology and effective masses
of SrFe$_2$P$_2$ which is the end member of the SrFe$_2$(As$_{1-x}$P$_x$)$_2$ series. We find that the hole and
electron sheets have quite different topology implying that a geometric nesting condition is not satisfied. This
perhaps, explains the lack of either superconductivity or magnetic order in this compound.  The many body mass
renormalisation are strongly sheet dependent and are much larger than expected from simple electron-phonon coupling.

\begin{figure}
\includegraphics[width = 7cm ]{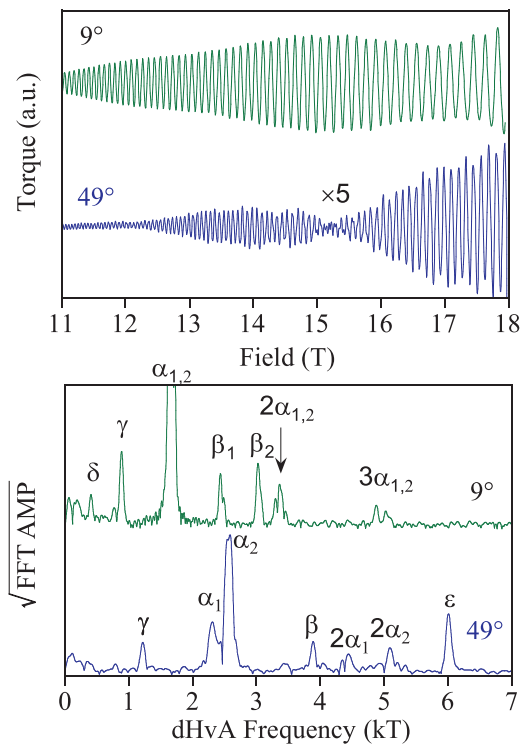}
\caption{dHvA oscillation data in SrFe$_2$P$_2$ at two different angles.(a) Raw torque signal with smooth polynomial
background subtracted (b) Fast Fourier transforms.  Note that the y-axis is the square-root of the FFT amplitude.}
\label{rawdata}
\end{figure}

\begin{figure*}
\includegraphics[width = 15cm]{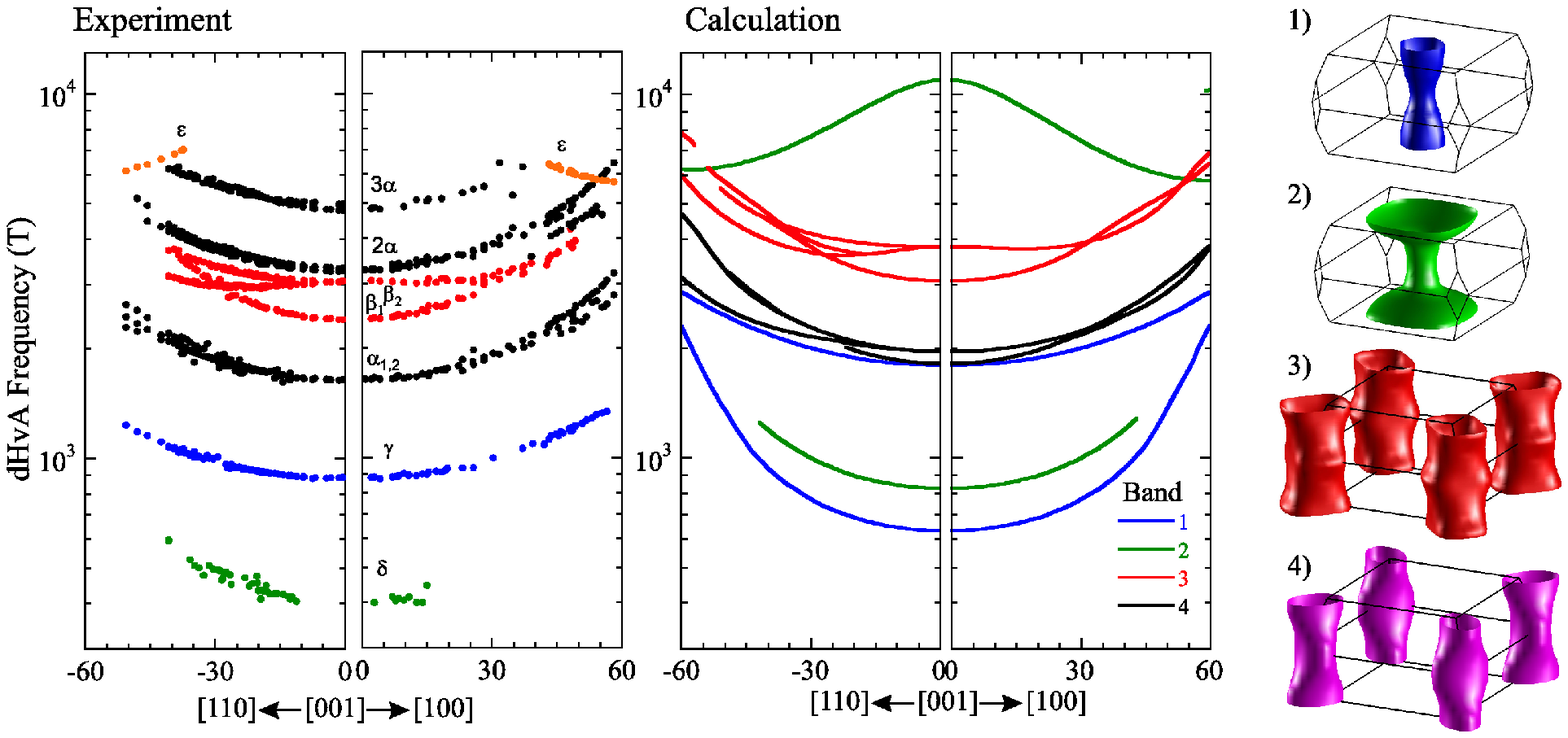}
\caption{(color online). Angle dependence of the dHvA frequencies for SrFe$_2$P$_2$.  The left panel
  of the angle plot shows the experimentally observed frequencies as
  the magnetic field is rotated from [110]$\rightarrow$ [001]
  $\rightarrow$ [100]. The right panel shows the corresponding
  predictions of the band structure calculation. At far
  right we show the Fermi surfaces associated with each band.} \label{angledep}
\end{figure*}

High quality single crystals of SrFe$_2$P$_2$ with residual resistivity ratios $\rho$(300\,K)/$\rho$(1.8\,K) greater
than 50, were grown from a Sn flux.  Torque magnetometry was performed using piezoresistive microcantilevers in high
fields \cite{EPAPS}. Band structure calculations were performed using an augmented plane wave plus local orbital method
as implemented in the WIEN2K code \cite{wien2k}. For these calculation we used the crystallographic parameters
$a=3.825$\,\AA, $c=11.612$\,\AA\, and $z_P=0.3521$ as determined by Mewis \cite{mewis}.

Fig.\ \ref{rawdata}(a) shows the raw torque signal for a single field sweep up to 18\,T. de Hass-van Alphen
oscillations are observed for fields above $\sim$ 4\,T. In Fig.\ \ref{rawdata} (b) we show the fast Fourier transform
(FFT) spectrum of the signal (in inverse field) measured at two angles $\theta=8^\circ$ and $\theta=49^\circ$, taken
relative to the $c$-axis of the crystal and rotating towards the [100] direction.  At all angles the spectrum is
dominated by a (split) single peak which we label $\alpha_{1,2}$, and in order to make the other smaller amplitude
peaks visible we show the square-root of the FFT amplitude in the figure.  In total seven frequencies are observed,
which we denote as $\alpha_1, \alpha_2, \beta_1, \beta_2, \gamma, \delta$ and $\epsilon$, and four others are harmonics
of $\alpha_{1,2}$. At higher angle additional splitting of these peaks is observed which likely comes from a small
$\sim1^\circ$ mosaic distribution of the crystal orientation. Note that the FFT peaks below $\sim 0.2$\,kT are noise
related.  Each of these frequencies are related to a extremal cross sectional area of the FS in momentum space $A_k$
via the Onsager relation F=$(\hbar/2\pi e)A_k$.

The calculated Fermi surface for SrFe$_2$P$_2$ consists of two concentric electron cylinders at Brillouin zone corner,
and two hole cylinders centered at the $\Gamma$ point (see Fig.\ \ref{angledep}) which is very similar to that
calculated for the \emph{non-magnetic} phase of SrFe$_2$As$_2$ \cite{sebastian}.  All of the sheets are significantly
warped.

\begin{figure}
\includegraphics[width = 6.2cm ]{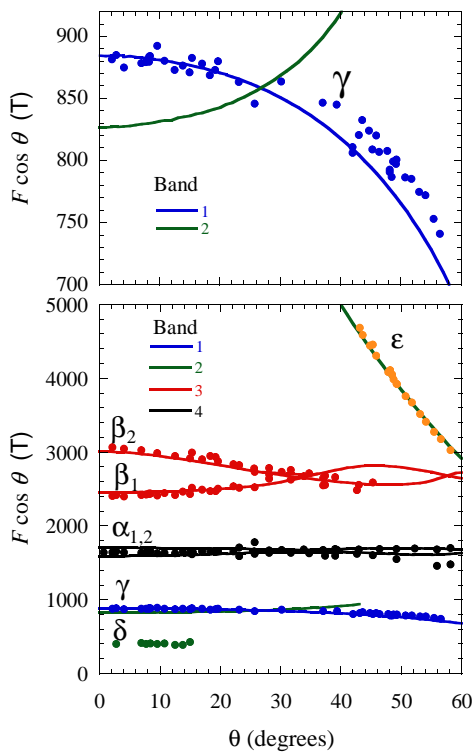}
\caption{(color online).Observed dHvA frequencies multiplied by $\cos \theta$ along with band structure predictions for
shifted energy bands.  The upper panel shows the data for the $\gamma$ orbit on an expanded scale } \label{fcostheta}
\end{figure}

Rotating away from the $c$-axis we can trace the full three-dimensional topology of the FS by looking at the evolution
of the dHvA frequencies as the cyclotron orbits traverse different parts of the FS. Figure \ref{angledep} (left) shows the
angle dependence of each of the observed frequencies with the field being rotated from [001] to [110] on the left and
[001] to [100] on the right. There is a high level of correspondence between the observed frequencies and the band
structure predictions shown in Fig.\ \ref{angledep} (right). The inner electron pocket (band 4) corresponds well to the
measured $\alpha$ frequencies, both in absolute cross sectional area (dHvA frequency) and angle dependence ($c$-axis
warping).  Similarly, band 3 matches the $\beta$ orbits.  Note the marked differences in angle dependence shown by the
$\beta$ orbits when rotated towards either the [100] or [110] directions. This is well reproduced in the calculations
and originates from the highly non-circular shape of this Fermi surface sheet (see Fig.\ \ref{angledep}).  Comparing in
detail the absolute frequencies of the $\alpha$ and $\beta$ with the calculations shows them both to be slightly
smaller.  A shift of the band energies up by 59\,meV and 49\,meV respectively for band 3 and 4 brings them into almost
perfect alignment (see Fig.\ \ref{fcostheta}).

\begin{figure}
\includegraphics[width =8cm]{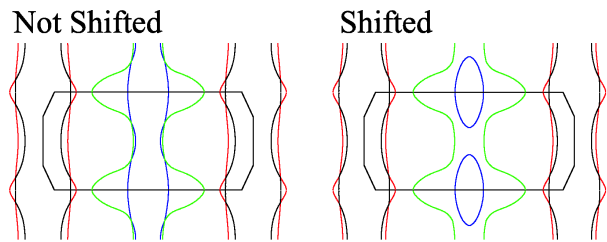}
\caption{(color online).Two dimensional sections through the FS in the (110) plane before and after shifting the bands
to agree with our dHvA measurements.  The Brillouin zone edges are marked.} \label{slices}
\end{figure}

The $\varepsilon$ orbit agrees very well with the maximum frequency from the large hole sheet (band 2), but is observed
over only a narrow range of angle (42--58$^\circ$).  The reason for this is apparent by looking at the slices through
the FS shown in Fig.\ \ref{slices}, where it can be seen that for this angle there is a large degree of nesting between
the warped sections of this band. Indeed, the bandstructure calculations show a large increase in the curvature factor
 $|\partial^2 A_k/\partial k^2_\||^{-\frac{1}{2}}$ (and hence signal amplitude) for these angles.

This leaves frequencies $\gamma$ and $\delta$ to assign. At first glance it appears as though the minima of either band
2 or band 1 could account for these. However, by plotting $F\cos\theta$ versus angle (see Fig.\ \ref{fcostheta}) it is
apparent that $\gamma$ cannot be a \emph{minimal} extremal crosssection because it decreases as the polar angle
increases. Hence, as shown in Fig.\ \ref{fcostheta} the minimum of band 2 has the wrong curvature to explain this
orbit. We therefore conclude that this orbit must originate from the maximum of band 1 as this is the only unassigned
orbit with the correct local topology.   For the frequencies to correspond we need to shift the energies of band 1 down
by 110\,meV. With this shift the curvature is in good agreement with data (see Fig.\ \ref{fcostheta}). Importantly, as
shown in Fig.\ \ref{slices} this shift causes the band 1 FS to be a closed three-dimensional ellipsoid. With this
assignment the remaining  $\delta$ orbit must come from the minimal (tubular section) of band 2. Only a small change in
the warping of this sheet would be needed to get complete agreement.

Fig.\ \ref{fcostheta} shows a comparison of all the observed frequencies with the shift band structure calculations and
Fig.\ \ref{slices} shows the resulting changes in the Fermi surface topology.  The shifts can be thought of as a fine
tuning of the bandstructure calculations.  For example, for band 3 the shift corresponds to a decrease in its volume by
0.044 electrons per units cell. A test of the consistency of the shift is to check for charge neutrality. SrFe$_2$P$_2$
is a compensated metal so the volumes of the electron and hole sheets should be exactly equal. With the shifts bands
1--4 contain -0.026, -0.284, 0.197, and  0.108 holes/electrons respectively, and so the charge balance remains almost
exact. The imbalance of 0.005 extra holes per unit cell is equivalent to $<2\%$ of the total volume of the hole sheets
showing the accuracy of our Fermi surface determination.

The effective masses, extracted \cite{EPAPS} by fitting the observed temperature dependent amplitude of the dHvA
oscillations to the conventional Lifshitz-Kosevich formula \cite{shoenberg}, are compared in Table I to the
corresponding bandstructure values.  The mass enhancements are strongly sheet dependent and range from
$\lambda=1-m^*/m_b) = 0.3$ for the smallest hole orbit to $\lambda=1.1$ for the inner electron sheet.  These
enhancements are similar to those observed in LaFePO \cite{coldea} and are much larger than expected from
electron-phonon coupling alone \cite{boeri}($\lambda_{ep}\simeq 0.25$). The table also shows the orbit specific dHvA
mean-free-paths. It is interesting to note that, as for LaFePO \cite{coldea}, the electron sheets have the longest
mean-free-paths and give the strongest dHvA signals. It is not clear to us if there is a fundamental reason for this.

The above illustrates that SrFe$_2$P$_2$\, has a FS which is highly dispersive in the $c$-axis and the electron and
hole pockets are far from fulfilling a nesting condition.  This may explain the observation of an almost isotropic
$H_{c2}$ in the analogous hole and electron doped superconducting 122 arsenides \cite{altarawneh,singleton}. In LaFePO,
the hole and electron FS sheets are much closer in shape and size \cite{coldea,sugawara} and this proximity to ($\pi,\pi$) nesting may well be a factor in determining why LaFePO superconducts while SrFe$_2$P$_2$ does not.

\begin{table}
\caption{Measured dHvA frequencies, effective masses ($m^*$) and mean free paths ($\ell$), along with the values from
the band structure calculations.  The experimental masses were determined at $\theta=9^\circ$ expect for the
$\varepsilon$ orbit which is at $\theta=49^\circ$.  The band structure values are all quoted at $\theta=0^\circ$ ($H
\|$ [001]). The final column show the ratio of measured effective mass to the band mass at the same angle (at
$\theta=49^\circ$ $m_b=1.98m_e$ for $2_{\rm max}$).}
\begin{tabular}{llllllll}
\hline \hline
\multicolumn{4}{c}{Experiment}&\multicolumn{4}{c}{Calculations}\\
 & F(kT) &$\frac{m^*}{m_e}$&$\ell$(nm)& Orbit & F(kT) &$\frac{m_b}{m_e}$ &$\frac{m^*}{m_b}$\\
\hline
            &           &               &   &$1_{\rm min}$  &0.632  &0.97&  \\
$\gamma$    & 0.89      & 1.49(2)       &58 &$1_{\rm max}$  &1.804  &1.07& 1.4\\
$\delta$    & 0.41      & 1.6(1)        &21 &$2_{\rm min}$  &0.828  &1.24& 1.3\\
$\epsilon$  & 6.02$^*$  & 3.41(5)$^*$   &90 &$2_{\rm max}$  &10.95  &2.30& 1.7\\
$\beta_1$   & 2.41      & 1.92(2)        &63&$3_{\rm min}$  &3.077  &1.25& 1.6\\
$\beta_2$   & 3.06      & 2.41(3)        &70&$3_{\rm max}$  &3.824  &1.70& 1.6\\
$\alpha_1$  & 1.637     & 1.13(1)       &100&$4_{\rm min}$  &1.823  &0.55& 2.1\\
$\alpha_2$  & 1.671     & 1.13(1)       &100&$4_{\rm max}$  &1.966  &0.60& 2.1\\
\hline
\end{tabular}
\end{table}

The differences with SrFe$_2$As$_2$ appear more complicated to answer because the phosphides and arsenides have much in
common in their local structure: the As and P ions are isoelectronic, they have a similar atomic radius (100pm and
115pm respectively), comparable ionization energies and coulomb interactions \cite{harrison_elementary_2004} and the
Pn-Pn distance is large compared with their respective molecular bond lengths (except for CaFe$_2$P$_2$, which is more
analogous to the collapsed tetragonal phase of CaFe$_2$As$_2$\cite{goldman_lattice_2008}). However, it is important to
notice that the 122 arsenides have Pn-Fe-Pn (Pn$\equiv$ pnictide) bond angles close to the ideal tetrahedral angle of
109.47$^\circ$. In the phosphides however, these bond angles are much larger, and for SrFe$_2$P$_2$ is 116.34$^\circ$
\cite{hoffmann_making_1985}.  The Fe-Pn distance is also smaller by $\sim 0.13$~\AA\, in the arsenides, leading to a
different bonding overlap. In particular, this results in differences in the calculated pnictogen density of
states\cite{gusteneu} which Yildirim has argued can dramatically alter the local magnetism of the Fe atoms
\cite{Yildirim2008a}.

The Pn-Fe-Pn bond angle departs from an ideal tetrahedral angle in both LaFePO (119.3$^\circ$) and SrFe$_2$P$_2$
(116.34$^\circ$) in a similar manner and this may be why neither of these compounds exhibit magnetic order. This
naturally accounts for the observed suppression of magnetism as SrFe$_2$As$_2$ is doped with P. However,
superconductivity arises at intermediate doping, and this implies that the magnetic order remains important. We
suggest that local interactions and a proximity to nesting work together to enhance $T_c$. The lower $T_c$ of the 122
compounds as compared to the 1111 compounds is then explained by the absence of nesting
\cite{Chubukov2008a,Mazin2008a,stanev_spin_2008,ran_nodal_2009}, and the phosphides have a much lower $T_c$ than the
arsenides because of the absence of local magnetic interactions \cite{xu,kivelson,Haule2008a,si_strong_2008}.

In conclusion, we have mapped out the full three dimensional Fermi surface of SrFe$_2$P$_2$, the isoelectronic sister
compound to the antiferromagnet SrFe$_2$As$_2$. The Fermi surface is in good overall agreement with the prediction of
our bandstructure calculations with small shifts in the band energies.  Unlike LaFePO the Fermi surface of
SrFe$_2$P$_2$ is far from fulfilling a geometric nesting condition, being composed of warped 2D electron cylinders, a
strongly warped outer hole cylinder and one closed hole pocket. The non-magnetically ordered, non-superconducting
ground state of SrFe$_2$P$_2$ is likely related to the combined absence of nesting and local moment interactions, compared to the analogous arsenides.

The authors would like to thank E. A. Yelland for technical assistance. Part of this work has been done with the financial support of EPSRC, Royal Society  and EU 6th Framework  contract
RII3-CT-2004-506239. Work at Stanford was supported by the U.S. DOE, Office of Basic Energy Sciences under contract
DE-AC02-76SF00515.


\end{document}